\documentclass[epsfig,12pt]{article}
\usepackage{graphicx}
\usepackage{epsfig}

\title{Difference of $\tilde \epsilon$ and $\epsilon$
in fitting the partameters of CKM matrix.}
\author{E.A.Andriyash \footnote{andriash@heron.itep.ru}, \\
{\small Moscow State University, Moscow, Russia} \\
G.G.Ovanesyan \footnote{ovanesyn@heron.itep.ru}, \\
{\small Moscow Institute of Physics and Technologies, Moscow, Russia} \\
M.I.Vysotsky \footnote{vysotsky@heron.itep.ru},\\
{\small ITEP, Moscow, Russia.}
}
\date{}
\begin{document}

\maketitle

\begin{abstract}

The difference between induced by box diagram value of $\tilde
\epsilon$ and experimentally measured value of $\epsilon$ is
estimated. It appeared to be around $5 - 10 \%$, depending on the
values of hadronic matrix elements. With this result, the fit of
CKM marix parameters within the SM is performed.
\end{abstract}

\newpage

\section{Introduction.}

It is well known that CP - violation in $K^0 - \bar K^0$ mixing is
described by the parameter $\tilde \epsilon$. Within the SM, this
parameter is given  by box diagrams. It depends in particular on
the CKM matrix elements. On the other hand, the experimentally
measured parameters are $\epsilon$ and $\epsilon'$. $\epsilon$ and
$\epsilon'$ enter the measured ratios of decay amplitudes of kaons
into $\pi \pi$ states. These amplitudes are superpositions of
amplitudes $A(K^0 \to (\pi \pi)_{I}) = A_I e^{i \delta_I}$ of kaon
decays into states with definite isospin $I=0,2$, $A_I$ are weak
amplitudes, $\delta_I$ are strong rescattering phases(see Appendix
for details). The parameter $\epsilon$ can be expressed as
\cite{Wolf}:

\begin{eqnarray}\label{1}
\epsilon = \tilde\epsilon + i \frac{Im A_0}{Re A_0}.
\end{eqnarray}

Within the SM, $Im A_0$ originates from the so-called strong
penguin diagrams. Amplitude $A_2$ also has an imaginary part which
originates from electro-weak penguin diagrams. That is why $Im A_0
>> Im A_2$. The ratio $\displaystyle{\frac{Im A_0}{Re A_0}}$ is
much smaller than $\tilde \epsilon$ and when the fit of the CKM
matrix parameters is performed, one equates the experimentally
measured value of $|\epsilon|$ and theoretical expression for
$|\tilde \epsilon|$, neglecting the term $\displaystyle{\frac{Im
A_0}{Re A_0}}$, see \cite{Laplace},\cite{Porter}. In particular it
was claimed in \cite{lectBur} that the contribution of
$\displaystyle{\frac{Im A_0}{Re A_0}}$ is "at most a 2\%
correction to $\epsilon$". The aim of the present paper is to take
this usually neglected term into account.

In order to estimate the ratio $\displaystyle{\frac{Im A_0}{Re
A_0}}$ we exploit the fact that it enters the expression for
$\displaystyle{\frac{\epsilon'}{\epsilon}}$\cite{Wolf}:

\begin{equation}\label{2}
\frac{\epsilon'}{\epsilon} = \frac{i}{\sqrt 2} e^{i(\delta_2
-\delta_0)} \frac{1}{\epsilon} \left[ \frac{Im A_2}{Re A_0} - w
\frac{Im A_0}{Re A_0} \right]~,
\end{equation}

\noindent where $\displaystyle{w = \frac{Re A_2}{Re A_0}}$.

The ratio $\displaystyle{\frac{\epsilon'}{\epsilon}}$ is
experimentally measured and great amount of work was done in order
to calculate it(see \cite{Gamiz} - \cite{Jamin} and refs.
therein). In particular the quantity $\displaystyle{\frac{Im
A_0}{Re A_0}}$ was computed theoretically using different methods.
We shall use the results of these computations.

We shall imply the following three step procedure for estimating
$\displaystyle{\frac{Im A_0}{Re A_0}}$.

At first step we neglect $Im A_0$. Then
$|\tilde\epsilon_{theor.}|$ coincides with $|\epsilon_{exp.}|$,
and we reproduce the results of \cite{Laplace},\cite{Porter}.

At second step we take into account that $Im A_0 \ne 0$, but
neglect the contribution of EW penguins in Eq.(\ref{2}). Then we
extract the value of $\displaystyle{\frac{Im A_0}{Re A_0}}$ from
experimentally measured quantity
$\displaystyle{\frac{\epsilon'}{\epsilon}}$ with the help of
Eq.(\ref{2}).

At third step we take into account the contribution of EW
penguins: $Im A_2 \ne 0$. The consequence is that one cannot
extract $\displaystyle{\frac{Im A_0}{Re A_0}}$ from Eq.(\ref{2}).
So one has to use the results of theoretical computation of
$\displaystyle{\frac{Im A_0}{Re A_0}}$.

Finally, we perform a fit of CKM matrix parameters, taking the
term $\displaystyle{\frac{Im A_0}{Re A_0}}$ in Eq.(\ref{1}) into
account and using numerical estimate of it, obtained at step 3.

\section{Difference between $\tilde \epsilon$ and $\epsilon$.}

The quantities $\epsilon$ and $\tilde \epsilon$ are related by
Eq.(\ref{1}). Taking into account that the phase of $\tilde
\epsilon$ is approximately $\frac{\pi}{4}$ \cite{Wolf} (see also
Appendix), from Eq.(\ref{1}) we deduce:

\begin{equation}
| \epsilon | = \left| \tilde \epsilon + i \frac{Im A_0}{Re
A_0}\right| = \sqrt{ \frac{1}{2} |\tilde \epsilon|^2 +
\left(\frac{1}{\sqrt{2}}|\tilde \epsilon| + \frac{Im A_0}{Re
A_0}\right)^2}.
\end{equation}

Thus:

\begin{equation}\label{3}
|\tilde \epsilon| = -\frac{1}{\sqrt{2}} \frac{Im A_0}{Re A_0} +
\sqrt{|\epsilon|^2 - \frac{1}{2} \left(\frac{Im A_0}{Re
A_0}\right)^2} \approx |\epsilon| - \frac{1}{\sqrt{2}} \frac{Im
A_0}{Re A_0}.
\end{equation}

The experimentally measured value is \cite{PDGW}:

\begin{equation}\label{33}
|\epsilon^{exp}| = 2.282(17) \times 10^{-3} .
\end{equation}

Now we start our procedure of estimating $|\tilde \epsilon|$.
\underline{At first step} we neglect $\displaystyle{\frac{Im
A_0}{Re A_0}}$ and obtain:

\begin{equation}\label{4}
|\tilde \epsilon| = | \epsilon^{exp}| = 2.282(17) \times 10^{-3}.
\end{equation}

This formula is always used in the fits of CKM matrix parameters,
see \cite{Laplace},\cite{Porter}.

\underline{Second step}: We take into account that $Im A_0 \ne 0$
but neglect $Im A_2$. Then Eq.(\ref{2}) reduces to:

\begin{equation}\label{5}
\frac{\epsilon'}{\epsilon} \approx -\frac{i}{\sqrt 2}
e^{i(\delta_2 -\delta_0 - \frac{\pi}{4})} \frac{w}{|\epsilon|}
\frac{Im A_0}{Re A_0}.
\end{equation}

Taking into account that $(\delta_0 - \delta_2)_{exp} = 42 \pm
4^{o}$ \cite{Chell}, we obtain the following expression for
$\displaystyle{\frac{Im A_0}{Re A_0}}$:

\begin{equation}\label{6}
\frac{Im A_0}{Re A_0} \approx - \frac{\sqrt 2 |\epsilon|}{w}
\frac{\epsilon'}{\epsilon}.
\end{equation}

Substituting experimental values from \cite{PDGW} we get:
\begin{eqnarray}\label{7}
&& \frac{\epsilon'}{\epsilon} = 1.8(4) \times 10^{-3}, \quad w =
0.045, \quad |\epsilon| = 2.282(17) \times 10^{-3} \Longrightarrow
\nonumber\\ && \frac{Im A_0}{Re A_0} = - (1.3 \pm 0.3) \times
10^{-4}.
\end{eqnarray}

In this way we get the following value of $|\tilde\epsilon|$,
which is the result of the second step:

\begin{equation}\label{8}
|\tilde \epsilon| = 2.37(2) \times 10^{-3}.
\end{equation}

This number coincides with the value obtained in \cite{lectVys},
Eqs.(9.3),(9.4).

\underline{Third step}: Now let us take into account the presence
of EW penguins: $Im A_2 \ne 0$. Then Eq.(\ref{2}) does not allow
to extract $\displaystyle{\frac{Im A_0}{Re A_0}}$ from the
experimental data and we need explicit theoretical result for
$\displaystyle{\frac{Im A_0}{Re A_0}}$. As announced in the
Introduction, such result was obtained in the literature while
calculating theoretically
$\displaystyle{\frac{\epsilon'}{\epsilon}}$.

In order to calculate $\displaystyle{\frac{\epsilon'}{\epsilon}}$
from Eq.(\ref{2}) , one needs theoretical expressions for $Im A_0$
and $Im A_2$ (the values of $Re A_0$, $Re A_2$, $|\epsilon|$,
$\delta_0 - \delta_2$ and $w$ are well measured experimentally).
Short review of the history of
$\displaystyle{\frac{\epsilon'}{\epsilon}}$ calculation can be
found in \cite{Jamin}. The expressions for $Im A_0$ and $Im A_2$
are usually presented in the following form:

\begin{eqnarray}\label{9}
&& Im A_0 = -\frac{ G_F}{\sqrt{2}} Im (V_{td} V_{ts}^*) P^{(0)}
(1-\Omega_{IB}), \nonumber\\ && Im A_2 = -\frac{ G_F}{\sqrt{2}} Im
(V_{td} V_{ts}^*) P^{(2)},
\end{eqnarray}

\noindent where

\begin{equation}
P^{(I)} = \sum_{i} y_i \langle Q_i\rangle_I, \quad I=0,2.
\end{equation}

Here $V_{td}$ and $V_{ts}^*$ are CKM matrix elements, $G_F$ -
Fermi constant, $\langle Q_i\rangle_{0,2}$ are matrix elements of
4-quark operators responsible for $K \rightarrow \pi \pi$ decays,
$y_i$ being their Wilson coefficients, $\Omega_{IB}$ introduces a
correction due to isospin breaking effects:
$\displaystyle{\Omega_{IB} = \frac{1}{w} \frac{(Im A_2)_{IB}}{Im
A_0}}$.

From (\ref{9}) we have:

\begin{equation}\label{10}
\frac{Im A_0}{Re A_0} = -\frac{ G_F}{\sqrt{2} Re A_0} Im (V_{td}
V_{ts}^*) P^{(0)} (1-\Omega_{IB}).
\end{equation}

This formula contains the CKM matrix elements (which we are going
to fit), but for the estimate of the small correction to $|\tilde
\epsilon|$ we can use mean values from \cite{PDGCKM}: $Im (V_{td}
V_{ts}^*) = 0.000127$. $Re A_0$ is well measured experimentally:
$Re A_0 = 3.33 \times 10^{-7} GeV$. Concerning $P^{(0)}
(1-\Omega_{IB})$, we use data from the calculations of
$\displaystyle{\frac{\epsilon'}{\epsilon}}$ done in \cite{Hambye},
which succeed in describing the experimental value of
$\displaystyle{\frac{\epsilon'}{\epsilon}}$.

Hadronic matrix elements were evaluated in \cite{Hambye} using
large $N_c$ - expansion. From Table 2 of \cite{Hambye} we find the
following range of values (corresponding to the quark condensate
value $\left(< \bar \psi \psi>\right)^{\frac{1}{3}} = 0.240 -
0.260 \, GeV$ at $\mu = 2 \, GeV $): $P^{(0)} (1 - \Omega_{IB}) =(
7.1 \pm 2.1) \times 10^{-2} GeV^3$.

Substituting this into (\ref{10}) we get: $\displaystyle{\frac{Im
A_0}{Re A_0}} = (-2.23 \pm 0.66) \times 10^{-4}$.

This leads to the following range of values for $|\tilde
\epsilon|$:

\begin{equation}\label{11}
2.39 \times 10^{-3} <\tilde \epsilon < 2.48 \times 10^{-3}.
\end{equation}

We have taken the paper \cite{Hambye} as an example, and similar
estimates can be made using other results, obtained in the
framework of $\displaystyle{\frac{\epsilon'}{\epsilon}}$
calculation (see \cite{Gamiz}-\cite{Jamin}).

The range of values for $|\tilde \epsilon|$ presented in
Eq.(\ref{11}) can be written as:\footnote{We note that a number, very close to our central value, can be extracted from \cite{Uli}. }

\begin{equation}\label{111}
|\tilde \epsilon| = (2.44 \pm 0.04) \times 10^{-3},
\end{equation}

\noindent and we use it in Section~\ref{Fit} to perform the fit of
the parameters of CKM matrix. As we see the value of
$|\tilde\epsilon|$ is larger than that obtained at step 1 by $(5 -
10)\%$.

\section{Fit of the parameters of CKM matrix }\label{Fit}

We use in our fit of the CKM matrix experimentally measured values
of modulus of matrix elements
$V_{ud}$,$V_{us}$,$V_{ub}$,$V_{cd}$,$V_{cs}$, $V_{cb}$ and also
$\tilde\epsilon$, $\Delta m_{B_d}$ and $sin 2\beta$.

We assume these experimentally measured data to be normally
distributed. Also the theoretical uncertainties are treated as
normally distributed. Let us note that other people treat
theoretical uncertainties in other way \cite{Laplace},
\cite{Porter}.

The most precise determination of $|V_{ud}|$ comes from the
averaging data from nuclear and neutron $\beta$ decays \cite{PDGCKM}:
\begin{equation}
|V_{ud}|=0.9734\pm 0.0008.
\end{equation}

From kaon semileptonic decays the element $|V_{us}|$ is determined
with the better accuracy than in other methods (like hyperon
semileptonic decays). We use the recent value \cite{PDGCKM}:
\begin{equation}
|V_{us}|=0.2196\pm 0.0026.
\end{equation}

From the inclusive and exclusive $B$-decays governed by the
transition $b\rightarrow u l^- {\bar{\nu}}_l$ we get \cite{PDGCKM}:

\begin{equation}
|V_{ub}|=0.0036\pm 0.0007.
\end{equation}

The element $|V_{cd}|$ was measured in deep inelastic scattering
of neutrinos and anti-neutrinos on nucleons with charm production \cite{PDGCKM}:
\begin{equation}
  |V_{cd}|=0.224\pm 0.016.
\end{equation}

The best accuracy in $|V_{cs}|$ comes from the measurement of the ratio of hadronic
$W$-decays to leptonic $W$-decays \cite{PDGCKM}:
\begin{equation}
|V_{cs}|=0.996\pm 0.013.
\end{equation}

The averaged value of $|V_{cb}|$ extracted from exclusive and
inclusive semileptonic $B$-decays including $c$ quark is \cite{PDGCKM}:
\begin{equation}
|V_{cb}|=0.041\pm 0.002.
\end{equation}

Theoretical expression for $|\tilde{\epsilon}|$ valid for $m_t >
m_W$ was first obtained in \cite{Vysold}. In modern notations it
looks like:
\begin{eqnarray}\label{epstilde}
  &&|\tilde{\epsilon}^{theo}|=\frac{G^2_F m^2_W m_K f^2_K}{12\sqrt{2}\pi^2\Delta m_K} B_K (\eta_{cc} S(x_c,x_c) Im[
  (V_{cs} V^*_{cd})^2]+\eta_{tt} S(x_t,x_t) Im[(V_{ts} V^*_{td})^2]\nonumber\\
  &&+2\eta_{ct} S(x_c,x_t)  Im[V_{cs} V^*_{cd} V_{ts} V^*_{td}] ).
\end{eqnarray}
Here, the $S(x_i, x_j)$ are usually called the Inami-Lim functions
\cite{inamilim}:
\begin{eqnarray}
  &&S(x)\equiv S(x_i,x_j)_{i=j}=x\left(\frac{1}{4}+\frac{9}{4(1-x)}-\frac{3}{2(1-x)^2}\right)-\frac{3}{2}\left
  (\frac{x}{1-x}\right)^3\ln{x},\nonumber\\
  &&S(x_i,x_j)_{i\neq j}=x_i x_j\left[\left(\frac{1}{4}+\frac{3}{2(1-x_i)}-\frac{3}{4(1-x_i)^2}\right)\frac{1}
  {x_i-x_j}\ln{x_i}+(x_i\leftrightarrow x_j)\right.\nonumber\\
  &&\left.-\frac{3}{4}\frac{1}{(1-x_i)(1-x_j)}\right],
\end{eqnarray}
where $x_i=m^2_i/m^2_W$ depend on the masses of $c$ quark and $t$
quark ($m_c=1.2\pm 0.2$ GeV\cite{PDG}, $m_t=174.3\pm 5.1$ GeV
\cite{PDG}, $m_W=80.42\pm 0.04$ GeV \cite{PDG}). The QCD
corrections have been calculated to next-to-leading order:
$\eta_{cc}=1.32\pm 0.32$ \cite{her}, $\eta_{tt}=0.574 \pm 0.01$
\cite{burasj}, $\eta_{ct}=0.47 \pm 0.04$ \cite{herr}. The kaon
decay constant  extracted from the $K^+\rightarrow \mu^+ \nu$ decay width equals:
$f_K=160.4\pm 1.9$ MeV \cite{PDG}. The
$K_S-K_L$ mass difference is $\Delta m_K=(3.491\pm 0.006)\times
10^{-15}$ GeV \cite{PDG}. The world average for the bag parameter
$B_K$ reads: $B_K=0.87\pm 0.06\pm 0.14_{quench}$ \cite{Lel}. Fermi
constant $G_F=1.16639(1)\times 10^{-5} GeV^{-2}$\cite{PDG}.

From the study of $B^0-\overline{B^0}$ oscillations the
experimental value of $|V_{td}|$ should be extracted:
\begin{equation}
  \Delta m_{B_d}=\frac{G^2_F}{6\pi^2}\eta_B m_{B_d} m^2_W S (x_t) f^2_{B_d}B_{B_d} |V_{td}V^*_{tb}|^2,
\end{equation}
where $\eta_B=0.55\pm 0.01$ \cite{burasj} is a QCD correction,
$m_{B_d}=5.2794\pm 0.0005$ GeV \cite{PDG} is the $B$ meson mass,
$m_W$ is the $W$ boson mass, $S(x_t)$ is the Inami-Lim function
for the box diagram, $x_t=\frac{m^2_t}{m^2_W}$, $f_{B_d}$ is the
$B$ meson decay constant, and $B_{B_d}$ is the so-called bag
factor. We use the following numerical value: $f_{B_d}
\sqrt{B_d}=230\pm 28 \pm 28$ MeV \cite{Bern}.

From $B$ decays to CP eigenstates containing charmonium and
neutral K-meson $\sin 2\beta$ is measured with good accuracy. The average result of Belle and BaBar
is \cite{sinbeta}:
\begin{equation}
 \sin 2\beta = 0.73 \pm 0.05
({\rm stat}) \pm 0.035 ({\rm syst}).
\end{equation}
Theoretical formula for $\sin 2\beta$ comes
from the consideration of the unitarity triangle:
\begin{equation}
  \sin 2\beta=\frac{2\bar{\eta}(1-\bar{\rho})}{\bar{\eta}^2+(1-\bar{\rho})^2}.
\end{equation}

The $\chi^2$ expression which we minimize looks like:
\begin{eqnarray}\label{chisq}
  &&\chi^2(A,\lambda,\rho,\eta)=\left(\frac{V^{theo}_{ud}-V^{exp}_{ud}}{\sigma_{V_{ud}}}\right)^2+\left(\frac{V^{theo}_{us}-V^{exp}_{us}}{\sigma_{V_{us}}}\right)^2+
  \left(\frac{V^{theo}_{ub}-V^{exp}_{ub}}{\sigma_{V_{ub}}}\right)^2 \nonumber\\
  &&+\left(\frac{V^{theo}_{cd}-V^{exp}_{cd}}{\sigma_{V_{cd}}}\right)^2+
  \left(\frac{V^{theo}_{cs}-V^{exp}_{cs}}{\sigma_{V_{cs}}}\right)^2+\left(\frac{V^{theo}_{cb}-V^{exp}_{cb}}{\sigma_{V_{cb}}}\right)^2\nonumber\\
  &&+\left(\frac{\Delta m^{theo}_{B_d}-\Delta m^{exp}_{B_d}}{\sigma_{\Delta_m}}\right)^2+\left(\frac{|\widetilde{\epsilon}^{theo}|-|\widetilde{\epsilon}^{exp}|}{\sigma_{\widetilde{\epsilon}}}\right)^2
  +\left(\frac{sin 2\beta^{theo}-sin 2\beta^{exp}}{\sigma_{sin2\beta}}\right)^2,\nonumber\\
\end{eqnarray}
where theoretical expressions depend on the Wolfenstein parameters
$A$, $\lambda$, $\rho$, $\eta$. Expression (\ref{chisq}) was
minimized varying $A$, $\lambda$, $\rho$, $\eta$.

Performing the fit we use the value of $|{\tilde\epsilon}^{theo}|$
from Eq.(\ref{111}). The main uncertainty in
$\tilde{\epsilon}^{theo}$ originates from that in $B_K$ and it
dominates in $\sigma_{\tilde{\epsilon}}$. That is why we use
$\sigma_{\tilde{\epsilon}}=0.4 \times 10^{-3}$.

Here are our results: $$ \lambda = 0.2229 \pm 0.0021$$ $$ A = 0.83
\pm 0.04$$ $$ \bar\eta = 0.35^{+0.05}_{-0.04}
$$ $$\bar\rho = 0.20^{+0.08}_{-0.09}$$ $$ \chi^2/n.d.o.f. =
8.1/5 \;\; .$$

In Fig.1 you can see a set of bounds on the parameters
$\bar{\rho}$ and $\bar{\eta}$ of the CKM matrix. They comprise three circles, two
branches of a hyperbola, and two straight lines. Three circles
originate from the $V_{ub}$  measurement (the green one), the measurement of $\Delta m_{B_d}$ (the red one)
 and from the lower bound on $\Delta m_{B_s}$ (the yellow one). The hyperbola originates from
the measurement of CP violation in the mixing of
$K$-mesons. Straight lines come from the
measurement of CP asymmetry in $B_d^0(\bar B_d^0) \to J/\Psi K$
decays.

\begin{figure}[!htb]
\centering
\epsfig{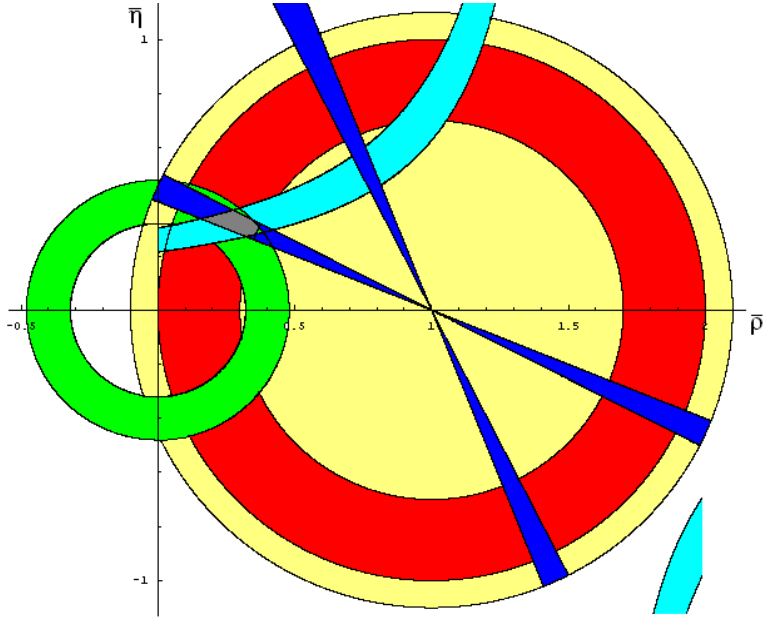} \caption{\em  The domains at $(\bar{\rho}, \bar{\eta})$ plane allowed at $1 \sigma$ from $V_{ub}$, $\Delta m_{B_d}$,
$\varepsilon_K$ and $\sin 2\beta$ measurements. 95{\rm \%}C.L. upper bound
from the search of $\Delta m_{B_s}$ is shown as well.}
\label{WW1Fermi}
\end{figure}

\section{Conclusions}\label{concl}

Numerical difference of the quantities $\epsilon$ and $\tilde \epsilon$
(which describe $CP$-violation in $K$-mesons) was estimated in the Standard Model.
Fit of CKM matrix patameters accounted for this difference was performed.

\section*{Acknowledgements}

We are grateful to Dr. Nierste for pointing our attention to reference \cite{Uli}.

This work was partially supported by FS NTP FYaF 40.052.1.1.1112 and by RFBR (grant N
00-15-96562). G.O. is grateful to Dynasty Foundation for partial support.

\appendix
\section{Basic formulas for $K^0$-$\bar{K}^0$ system}\label{epsilon}

It is known that states $K^0$ and $\bar K^0$ are not mass
eigenstates. Mass eigenstates are their linear combinations:

\begin{eqnarray}\label{Mstates}
&& K_+ = \frac{1}{\sqrt{1+ \mid \tilde\varepsilon \mid^2}}
\left[\frac{K^0 +\bar K^0}{\sqrt{2}} + \tilde\varepsilon \frac{K^0
- \bar K^0}{\sqrt{2}} \right], \nonumber\\ && K_-
=\frac{1}{\sqrt{1+ \mid \tilde\varepsilon \mid^2}} \left[\frac{K^0
-\bar K^0}{\sqrt{2}} + \tilde\varepsilon \frac{K^0 +\bar
K^0}{\sqrt{2}} \right].
\end{eqnarray}

Let's denote matrix elements of the effective Hamiltonian between
$K^0$ and $\bar K^0$ states as follows:

\begin{eqnarray}
&& <K^0 \mid H\mid K^0 > = <\bar K^0 \mid H \mid \bar K^0
> =  M -\frac{i}{2}\Gamma, \nonumber\\ && <K^0 \mid H \mid \bar K^0
> =  M_{12} - \frac{i}{2} \Gamma_{12}, \nonumber\\
&& <\bar K^0 \mid H \mid  K^0
> =  M_{12}^* - \frac{i}{2} \Gamma_{12}^*.
\end{eqnarray}

The eigenvalues and eigenvectors of this matrix Hamiltonian are:

\begin{eqnarray}
&& \lambda_{\pm} = M -\frac{i}{2}\Gamma \pm \sqrt{(M_{12} -
\frac{i}{2}\Gamma_{12})(M_{12}^* - \frac{i}{2}\Gamma_{12}^*)},
\nonumber\\ &&  \left\{ \begin{array}{l} M_+ = pM^0 + q \bar M^0
\\ M_- = pM^0 - q \bar M^0 \end{array} \right. \;\; , \;\;
\frac{q}{p} = \sqrt{\frac{M_{12}^* -
\frac{i}{2}\Gamma_{12}^*}{M_{12} -\frac{i}{2}\Gamma_{12}}}.
\end{eqnarray}

Introducing quantity $\tilde\varepsilon$ according to the
following definition:
\begin{equation}\label{eps}
\frac{q}{p} = \frac{1-\tilde\varepsilon}{1+\tilde\varepsilon},
\end{equation}

we come to Eq.(\ref{Mstates}).

Taking into account that $\Gamma_{12}$ is real and $\displaystyle{
\frac{Im M_{12}}{Re M_{12}}} \sim 0.1$ \cite{lectVys} we get the following
expression:

\begin{eqnarray}\label{qp}
\frac{q}{p} \approx 1 - \frac{i Im M_{12}}{M_{12} - \frac{i}{2} \Gamma_{12}}.
\end{eqnarray}

Eigenvalues of Hamiltonian may be written as $\lambda_{\pm} =
(m_{\pm} - \frac{i}{2}\Gamma_{\pm})^2$, where $m_{\pm}$ are masses
of corresponding states and $\Gamma_{\pm}$ - their widths. Then
denoting $K_+$ and $K_-$ states as $K_S$ and $K_L$ respectively,
we have  $\lambda_--\lambda_+ = 2 m_K (m_L - m_S - \frac{i}{2}
(\Gamma_L - \Gamma_S))$. On the other hand  $\lambda_--\lambda_+ =
- 2 \sqrt{(M_{12} - \frac{i}{2}\Gamma_{12})(M_{12}^* -
\frac{i}{2}\Gamma_{12}^*)} \sim - 2 (M_{12} - \frac{i}{2} \Gamma_{12}).$
This leads to:

\begin{eqnarray}
\frac{q}{p} \approx 1 + \frac{i Im M_{12}/m_K}{(m_L - m_S - \frac{i}{2}
(\Gamma_L - \Gamma_S))} \approx 1-2 \tilde \epsilon.
\end{eqnarray}

Taking into account that $\Gamma_S << \Gamma_L$ and $\Delta
m_{LS} \approx \Gamma_S/2$, we obtain:

\begin{eqnarray}
\tilde \epsilon \approx -\frac{i Im M_{12}/2 m_K}{(\Delta m_{LS} - \frac{i}{2}
(\Gamma_L - \Gamma_S))} \approx e^{-i \frac{3 \pi}{4}} \frac{Im M_{12}/2
m_K}{\sqrt{2} \Delta m_{LS}}.
\end{eqnarray}

Thus calculating $Im M_{12}$ within the SM we find the theoretical
prediction for $\tilde \epsilon$. (Let us note that since $Im M_{12}$ is negative,
the phase of $\tilde \epsilon$ approximately equals $\frac{\pi}{4}$).

Now we proceed to decays of kaons into pairs of pions, whose
amplitudes are well measured experimentally.

It is convenient to deal with the amplitudes of the decays into
the states with definite isospin:

\begin{eqnarray}
&& A(K^0 \to \pi^+ \pi^-) =
\frac{a_2}{\sqrt 3} e^{i\xi_2}e^{i\delta_2} + \frac{a_0}{\sqrt 3}
\sqrt{2} e^{i\xi_0} e^{i\delta_0} \nonumber\\
&& A(\bar K^0 \to \pi^+ \pi^-) = \frac{a_2}{\sqrt 3}
e^{-i\xi_2}e^{i\delta_2} + \frac{a_0}{\sqrt 3} \sqrt{2}
e^{-i\xi_0} e^{i\delta_0} \nonumber\\
&& A(K^0 \to \pi^0 \pi^0) = \sqrt{\frac{2}{3}} a_2 e^{i\xi_2}e^{i\delta_2} -
\frac{a_0}{\sqrt 3} e^{i\xi_0} e^{i\delta_0} \nonumber\\
&& A(\bar K^0 \to \pi^0 \pi^0) = \sqrt{\frac{2}{3}} a_2
e^{-i\xi_2}e^{i\delta_2} - \frac{a_0}{\sqrt 3} e^{-i\xi_0}
e^{i\delta_0}
\end{eqnarray}

where ``2'' and ``0'' are the
values of ($\pi\pi$) isospin, $\xi_{2,0}$ are the (small) weak
phases which originate from CKM matrix and $\delta_{2,0}$ are the
strong phases of $\pi\pi$-rescattering.

Experimentally measured quantities are:

\begin{eqnarray}
&& \eta_{+-} = \frac{A(K_L \to\pi^+\pi^-)}{A(K_S \to\pi^+ \pi^-)},
\nonumber\\ &&  \eta_{00} = \frac{A(K_L\to\pi^0\pi^0)}{A(K_S \to
\pi^0 \pi^0)}.
\end{eqnarray}

For the amplitudes of $K_L$ and $K_S$ decays into $\pi^+ \pi^-$ we
obtain:

\begin{eqnarray}
&& A(K_L \to\pi^+ \pi^-) = \frac{1}{\sqrt 2} \left[
\frac{a_2}{\sqrt 3} e^{i\delta_2} 2i\sin\xi_2 + \frac{a_0}{\sqrt
3} \sqrt 2 e^{i\delta_0} 2i\sin\xi_0 \right] + \nonumber\\ && +
\frac{\tilde \epsilon}{\sqrt 2} \left[ \frac{a_2}{\sqrt 3}
e^{i\delta_2} 2\cos\xi_2 + \frac{a_0}{\sqrt 3} \sqrt 2
e^{i\delta_0} 2\cos\xi_0 \right] ,\nonumber\\ && A(K_S \to\pi^+
\pi^-) = \frac{1}{\sqrt 2} \left[ \frac{a_2}{\sqrt 3}
e^{i\delta_2} 2\cos\xi_2 + \frac{a_0}{\sqrt 3} \sqrt 2
e^{i\delta_0} 2\cos\xi_0 \right],
\end{eqnarray}

\noindent where in the last equation we omit the terms which are
proportional to the product of two small factors, $\tilde
\epsilon$ and $\sin\xi_{0,2}$. For the ratio of these amplitudes
we get: $$ \eta_{+-} = \frac{A(K_L \to\pi^+\pi^-)}{A(K_S \to\pi^+
\pi^-)} = \tilde \epsilon + i\frac{\sin\xi_0}{\cos\xi_0} +\frac{i
e^{i(\delta_2 - \delta_0)}}{\sqrt 2} \frac{a_2 \cos\xi_2}{a_0
\cos\xi_0} \left[ \frac{\sin\xi_2}{\cos\xi_2} -
\frac{\sin\xi_0}{\cos\xi_0}\right],$$ where we neglect the terms
of the order of $(a_2/a_0)^2 \sin\xi_{0,2}$, because $a_2/a_0
\approx 1/22$.

The analogous treatment of $K_{L,S} \to \pi^0 \pi^0$ decay
amplitudes leads to: $$ \eta_{00} = \frac{A(K_L
\to\pi^0\pi^0)}{A(K_S \to \pi^0 \pi^0)} = \tilde \epsilon +
i\frac{\sin\xi_0}{\cos\xi_0} - ie^{i(\delta_2 - \delta_0)} \sqrt 2
\frac{a_2 \cos\xi_2}{a_0 \cos\xi_0}
\left[\frac{\sin\xi_2}{\cos\xi_2} - \frac{\sin\xi_0}{\cos\xi_0}
\right].
$$

Introducing conventional quantities $\epsilon = \frac{2}{3}
\eta_{+-} + \frac{1}{3} \eta_{00}$ and $\epsilon' =
\frac{1}{3}\eta_{+-} - \frac{1}{3} \eta_{00}$, we get:

\begin{eqnarray}\label{29}
&& \epsilon' = \frac{i}{\sqrt 2} e^{i(\delta_2 -\delta_0)}
\frac{1}{Re A_0} \left[ Im A_2 - w Im A_0 \right]\nonumber\\ &&
\epsilon = \tilde \epsilon + i \frac{Im A_0}{Re A_0},
\end{eqnarray}

where $A_{2,0} \equiv e^{i\xi_{2,0}} a_{2,0}$, and
$\displaystyle{w = \frac{Re A_2}{Re A_0} \approx
\frac{a_2}{a_0}}$.

Equations (\ref{29}) are our starting point in the present paper;
see Introduction.

\end{document}